\documentclass[prd,twocolumn,preprintnumbers]{revtex4}

\usepackage{rotate}
\usepackage{epsfig}

\newcommand{\fpi}{f_\pi}
\newcommand{\mpi}{m_\pi}
\newcommand{\gev}{\, {\rm GeV}}
\newcommand{\mev}{\, {\rm MeV}}
\newcommand{\fm}{\, {\rm fm}}
\newcommand{\non}{\nonumber}

\newcommand{\be}{\begin{equation}}
\newcommand{\ee}{\end{equation}}
\newcommand{\bea}{\begin{eqnarray}}
\newcommand{\eea}{\end{eqnarray}}
\newcommand{\ra}{\rightarrow}
\newcommand{\chpt}{$\chi$PT}

\newcommand{\smallfigsize}{\columnwidth}
\newcommand{\myfigsize}{\columnwidth}

\begin{document}

\title{Convergence of Chiral Effective Field Theory}

\author{R.~D.~Young, D.~B.~Leinweber and A.~W.~Thomas}
\affiliation{    Special Research Centre for the
                 Subatomic Structure of Matter\\
                 and Department of Physics and Mathematical Physics,
                 University of Adelaide, Adelaide SA 5005,
                 Australia}

\begin{abstract}
We formulate the expansion for the mass of the nucleon as
a function of pion mass within chiral perturbation theory using a
number of different ultra-violet regularisation schemes; including
dimensional regularisation and various finite-ranged
regulators. Leading and next-to-leading order non-analytic
contributions are included through the standard one-loop Feynman
graphs. In addition to the physical nucleon mass, the expansion is
constrained by recent, extremely accurate, lattice QCD data obtained
with two flavors of dynamical quarks. The extent to which different
regulators can describe the chiral expansion is examined, while
varying the range of quark mass over which the expansions are
matched. Renormalised chiral expansion parameters are recovered from
each regularisation prescription and compared. We find that the
finite-range regulators produce consistent, model-independent results
over a wide range of quark mass sufficient to solve the chiral
extrapolation problem in lattice QCD.
\end{abstract}

\preprint{ADP-02-104/T542}

\maketitle


\section{Introduction}
The importance of incorporating the chiral non-analytic behavior of
hadronic observables in the extrapolation of lattice QCD simulation
results has been highlighted in numerous studies over the past few
years [1--9].
In an attempt to avoid model dependence, early lattice QCD
extrapolations considered simple polynomial functions of the pion
mass.  However it is now widely accepted that the model-independent, 
chiral non-analytic behavior predicted by chiral perturbation
theory ($\chi$PT) must be incorporated in any quark mass extrapolation
function in order to preserve the properties of QCD.

What is more controversial is the manner in which the effective field
theory is regulated.  Historically, most formulations of \chpt\ are
based on dimensional regularisation and surprisingly other
regularisation schemes are often regarded as models.  However, the
physical predictions of the effective field theory must be
regularisation and renormalisation scheme independent, such that other
schemes are possible and may provide advantages over the traditional
approach.  Indeed, Donoghue {\it et al.}~\cite{Donoghue:1998bs} have
already reported the improved convergence properties of effective
field theory formulated with what they call a ``long-distance
regulator,'' such as a dipole or Gaussian form, designed to suppress
large three-momentum contributions to the loop integrals of
heavy-baryon \chpt.

The origin of the controversy lies in the phenomenological
interpretation of the long-distance regulator as a form factor
describing the finite size of the meson-cloud source. In this light,
it is common to think of the regulator as a model, removing the
possibility of systematic improvement through the calculation of
higher order terms. However, the interpretation of the regulator as a
physical form factor is valid only when the terms analytic in quark
mass --- the terms involving even powers of $m_\pi$ in
Eq.(\ref{eq:eftexp}) below --- are either set to zero or determined
independent of data by some model of hadron structure. While terms are
not constrained by the effective field theory, one can maintain
model-independence by including them as parameters fit to data, order
by order in the expansion. However there is an obvious corollary of
this discussion. To suppress the size of the contributions of the
analytic terms, one should consider the use of a phenomenologically
motivated regulator. The suppression of the higher-order analytic
terms in the corresponding residual expansion that may result from a
good choice holds the promise of improved convergence properties in
that expansion.

It is extremely important to explore this possibility of improved
convergence.  At present it is clear that to link full
dynamical-fermion lattice QCD simulation results for hadron masses
with dimensionally-regulated $\chi$PT, new simulation results are
needed in the range from the physical pion mass to 0.1 GeV${}^2$
\cite{Bernard:2002yk}; a daunting task even with the multi-Teraflops
resources available over the next five years.  Perhaps even more
important is the formal issue of the formulation of $SU(3)$-flavor
chiral perturbation theory, associated with a strange-quark mass,
which likely lies beyond the applicable range of
dimensionally-regulated chiral perturbation theory for most baryon
observables.

This study will quantitatively compare the chiral expansion of the
nucleon mass for six different regularisation schemes, including
traditional dimensional regularisation and a variety of long distance
regulators.  Our focus will be to determine over what range in quark
mass (or squared pion mass) {\it all} chiral expansions agree at the
1\% level.  It will become apparent that the long distance regulators
do indeed have dramatically improved convergence properties and can
imitate the dimensionally regulated chiral expansion over a wider
range than the converse.  This range will also be reported.  Finally,
we will discuss the physics, explaining why dimensional regularisation
is a poor choice of regulator for effective field theory, in the
context of lattice extrapolations.  On the other hand, we reach the
exciting conclusion that effective field theory formulated with a
smooth long-distance regulator provides model-independent chiral
expansions agreeing at the 1\% level over a range exceeding $0 \le
m_\pi^2 \le 0.8$ GeV$^2$.  Thus the chiral extrapolation problem has
been solved.


\section{Chiral Effective Field Theory}
Chiral perturbation theory (\chpt) is a low-energy effective field
theory for QCD. Using this effective field theory, the low-energy
properties of hadrons can be expanded about the limit of vanishing
momenta and quark mass. In particular, in the context of the
extrapolation of lattice data, \chpt\ provides a functional form
applicable as $m_q$ approaches zero or, equivalently, $\mpi\to 0$
through the Gell-Mann--Oakes--Renner (GOR) relation $\mpi^2\propto
m_q$ \cite{Gell-Mann:1968rz}. Goldstone boson loops play
an especially important role in the theory as they give rise to
non-analytic behaviour as a function of quark mass. The low-order,
non-analytic contributions arise from the pole in the Goldstone boson
propagator and hence are {\em model-independent} \cite{Li:1971vr}.
However, because the analytic variation of hadron properties is not
constrained by chiral symmetry, the expansions contain free parameters
which must be determined empirically by comparison with data.

Effective field theory then tells us that the formal expansion of the
nucleon mass about the SU(2) chiral limit is:
\bea
M_N &=& a_0 + a_2 m_\pi^2 + a_4 m_\pi^4 \non\\
& &     + \sigma_{{\rm N} \pi}(m_\pi, \Lambda) 
        + \sigma_{\Delta \pi}(m_\pi, \Lambda)+\ldots\, ,
\label{eq:eftexp}
\eea
where $\sigma_{{\rm B} \pi}$ is the self-energy arising from a $B\pi$
loop (with $B$ = $N$ or $\Delta$) and $\Lambda$ is a parameter
associated with the regularisation. These $N$ and $\Delta$ loops
generate the leading and next-to-leading non-analytic (LNA and NLNA)
behaviour, respectively. The expansion has deliberately been written
in this form to highlight that the theory is equivalently defined for
an arbitrary regulator. This equivalence has been demonstrated by
Donoghue {\it et.~al.}~\cite{Donoghue:1998bs}, where the convergence
properties of SU(3) \chpt\ were improved by implementing a
finite-range regulator (FRR), which they term ``long-distance
regularisation''. More recently this work has been extended to
include decuplet baryons \cite{Borasoy:2002jv}.

The traditional approach within the literature is to use dimensional
regularisation to evaluate the self-energy integrals. Under such a
scheme the $NN\pi$ contribution simply becomes $\sigma_{{\rm N}
\pi}(\mpi,\Lambda)\to c_{\rm LNA}\mpi^3$ and the analytic terms,
$a_n\mpi^n$ (with $n$ even), undergo an infinite
renormalisation. The $\Delta$ contribution produces a logarithm, so
that the complete series expansion of the nucleon mass about $\mpi= 0$
is:
\bea
M_N &=& c_0 + c_2 m_\pi^2 + c_4 m_\pi^4 \non\\
& &   + c_{\rm LNA} m_\pi^3 
      + c_{\rm NLNA} m_\pi^4 \ln m_\pi + \ldots\, ,
\label{eq:drexp}
\eea
where the $a_i$ have been replaced by the renormalised (and
finite) parameters $c_i$.

As mentioned above, coefficients of low-order non-analytic
contributions are known, with~\cite{Gasser:1988rb,Lebed:1994yu}
\bea
c_{\rm LNA}&=&-\frac{3}{32\,\pi\,\fpi^2}\,g_A^2\, , \non\\
c_{\rm NLNA}&=&\frac{3}{32\,\pi\,\fpi^2}\,\frac{32}{25}g_A^2\,
\frac{3}{4\,\pi\,\Delta}\, .
\eea
Although strictly one should use values in the chiral limit, we take
the experimental numbers with $g_A=1.26$, $\fpi=0.093\gev$, the
nucleon--delta mass splitting, $\Delta=0.292\gev$
and the mass scale associated with the logarithms will be taken to be
1 GeV (i.e. essentially $4 \pi f_\pi$).

One would expect that it would be possible to reach the physical pion
mass from the chiral limit with this expansion truncated beyond $c_6
\mpi^6$. Unfortunately, the convergence of such an expansion seems to
break down fairly quickly at higher pion masses. It is also important
to realise that Eq. (\ref{eq:drexp}) was derived in the limit
$\mpi/\Delta \ll 1$. At just twice the physical pion mass this ratio
approaches unity. Mathematically the region $m_\pi \approx \Delta$ is
dominated by a square root branch cut which starts at $m_\pi =
\Delta$. Using dimensional regularisation this takes the form
\cite{Banerjee:1996wz}:
\bea
\frac{6 g_A^2}{25 \pi^2 f_\pi^2} \bigg[& (\Delta^2 - m_\pi^2)^{\frac{3}{2}} 
\ln (\Delta + m_\pi -\sqrt{\Delta^2 - m_\pi^2}) \non\\
& - \frac{\Delta}{2}(2 \Delta^2 - 3 m_\pi^2) \ln m_\pi \bigg],
\label{eq:log}
\eea
for $m_\pi < \Delta$, while for $m_\pi > \Delta$ the first logarithm
becomes an arctangent. Clearly, to access the higher quark masses in the
chiral expansion, currently of most relevance to lattice QCD,
one requires a more sophisticated expression than
that given by Eq.~(\ref{eq:drexp}).

Even ignoring the $\Delta \pi$ cut for a short time, the formal
expansion of the $N\, \ra\, N\,\pi\, \ra\, N$ self-energy integral,
$\sigma_{{\rm N} \pi}$, has been shown to have poor convergence
properties. Using a sharp, ultra-violet cut-off, Wright showed
\cite{SVW} that the series expansion, truncated at ${\cal O}(\mpi^4)$,
diverged for $m_\pi > 0.4$ GeV. This already indicated that the series
expansion motivated by dimensional regularisation would have a slow
rate of convergence.

The main issue of the convergence of this truncated series,
Eq.~(\ref{eq:drexp}), is linked to the formalism in which it is
derived from the general form of Eq.~(\ref{eq:eftexp}). The
dimensionally regulated approach requires that the pion mass should
remain much lighter than every other mass scale involved in the
problem. This requires that $\mpi/\Lambda_{\chi {\rm SB}}\ll 1$ and
$\mpi/\Delta\ll 1$. An additional scale, mentioned briefly above, is
set by the physical extent of the source of the pion field. This
scale, which is of order $R_{SOURCE}^{-1}$, corresponds to the
transition between the rapid, non-linear variation required by chiral
symmetry and the smooth, {\em constituent-quark} like mass behaviour
observed in lattice simulations at larger quark mass
\cite{Cloet:2002eg}. An alternative to dimensional regularisation
would be to regulate Eq.~(\ref{eq:eftexp}) with a finite ultra-violet
cut-off (mass parameter $\Lambda$), which physically corresponds to
the fact that the source of the meson cloud is an extended structure.

In summary, while the low-energy effective field theory can be very
useful in describing the quark mass dependence of hadron properties at
very low values of $m_\pi$, its utility in the context of lattice QCD
appears to have been limited by the tendency in the literature to
focus on a single type of regulator (i.e. dimensional regularisation).
We now investigate other possible regularisation schemes in order to
see whether they are able to ameliorate the problem.


\section{Constraining the Chiral Expansion}
\label{sec:constrain}
Here we investigate the effective rate of convergence of the chiral
expansions obtained using different functional forms for the
regulator, together with an analysis of the dimensionally regulated
approach. To analyse the merits of various regularisation schemes, at
best, we would require exact knowledge of how the nucleon mass varies
with quark mass. Having only one value for the experimental nucleon
mass we cannot determine the parameters that govern the quark mass
dependence of $M_N$ without taking some information from alternate
sources.

Lattice QCD provides a non-perturbative method for studying the
variation of $M_N$ with $\mpi$. In principle, this allows one to fix
the parameters of the chiral expansion using data obtained in
simulations performed at varying quark mass. Lattice simulations of
full QCD are restricted to the use of relatively heavy quarks and
hence it is not clear, a priori, whether the effective field theory
expansion is capable of linking to even the lightest simulated quark
mass where $\mpi\sim 500\mev$. An analysis of each regularisation
scheme is necessary to determine the effective range of applicability.

Taking as input the physical nucleon mass and the latest lattice QCD
results of the CP-PACS Collaboration \cite{AliKhan:2001tx} enables us
to constrain an expression for $M_N$ as a function of the quark
mass. The lattice results have been obtained using improved gluon and
quark actions on fine, large volume lattices with high
statistics\footnote{Simulations are performed using an Iwasaki gluon
action \cite{Iwasaki:1985we} and the mean-field improved clover
fermion action.}. In this work we concentrate on only those results
with $m_{\rm sea} = m_{\rm val}$ and the two largest values of $\beta$
(i.e.  the finest lattice spacings,\footnote{Note that we employ the
UKQCD method \cite{Allton:1998gi} to set the physical scale, for each
quark mass, via the Sommer scale $r_0=0.5\fm$
\cite{Sommer:1994ce,Edwards:1998xf}. This choice is ideal in the
present context because the static quark potential is insensitive to
chiral physics.} $a\sim 0.09$ -- $0.13\fm$). This ensures that the
results obtained represent accurate estimates of the continuum,
infinite-volume theory at the simulated quark masses. The lattice data
lies in the intermediate mass region, with $\mpi^2$ between $0.3$ and
$0.7\gev^2$. One may ask whether the effective field theory has any
applicability at this mass scale. The following analysis will answer
this question in a model-independent fashion.

In order to remove the bias of choosing any particular regularisation
scheme we allow each scheme to serve as a constraint curve for the
other methods. In this way we generate six (one for each
regularisation) different constraint curves that describe the quark
mass dependence of $M_N$.  The first case corresponds to the truncated
power series obtained through the dimensionally regulated (DR)
approach, Eq.~(\ref{eq:drexp}). We work to analytic order $\mpi^6$,
which is necessary in order to counter the large, non-analytic
contribution arising at order $\mpi^4\log\mpi$. The second procedure
(labeled ``BP'') takes a similar form but the branch point obtained
from the dimensionally regularised $N\to\Delta\pi$ transition is
retained in its full functional form.  That is, the logarithm in
Eq.~(\ref{eq:drexp}) is replaced by the full expression,
Eq.~(\ref{eq:log}), which ensures the correct non-analytic structure
where the logarithm converts to an arctangent above the branch
point. We refer to this form as the dimensionally-regulated
branch-point (BP) approach. Finally, we use four different functional
forms for the finite-ranged, ultra-violet vertex regulator. These are
namely the sharp-cutoff (SC), $\theta(\Lambda-k)$; monopole (MON),
$\Lambda^2/(\Lambda^2+k^2)$; dipole (DIP),
$\Lambda^4/(\Lambda^2+k^2)^2$; and Gaussian (GAU),
$\exp(-k^2/\Lambda^2)$. It is commonly assumed in the literature that
the functional form chosen for the finite-range regulator will
introduce model dependence \cite{Becher:1999he} and these rather
different forms are chosen in order to test this hypothesis.

Since, in the case of finite-range regularisation, we allow the
regulator parameter to be tuned to the lattice QCD data points, we do
not require an analytic term at order $\mpi^6$. The use of a
finite-ranged regulator will implicitly include a term of order
$\mpi^6$, together with a string of higher order terms. Tuning the
regulator parameter optimises the efficiency of the one-loop chiral
expansion. In particular, it naturally suppresses the large
non-analytic contributions which would otherwise dominate at high pion
masses. This is in contrast to the dimensional regulated case, where
the inclusion of the term in $\mpi^6$ is a practical necessity because 
of the large, non-analytic contribution at order $\mpi^4 \ln\mpi$.

\begin{figure}[t]
\begin{center}
{\epsfig{file=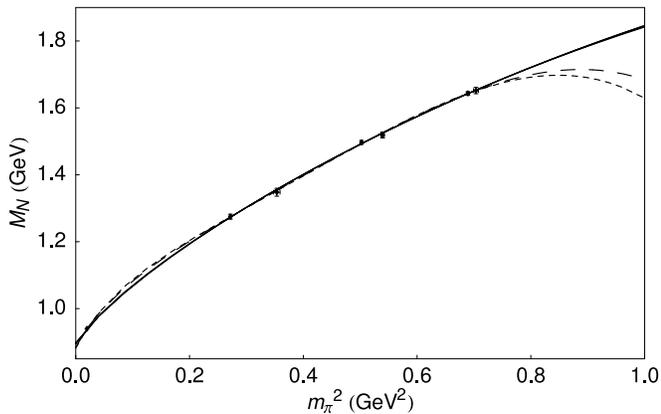, width=\columnwidth}}
\caption{Various regularisation schemes providing constraint curves
for the variation of $M_N$ with pion mass. The short dash curve
corresponds to the simple dimensional regularisation scheme (DR) and
the long dash curve to the more sophisticated dimensionally regulated
approach (BP), which keeps the correct non-analytic structure at the
$\Delta \pi$ branch point. The four finite-range regulators (solid
curves) are indistinguishable at this scale. Oscillations of the
dimensionally regulated schemes (dashed curves) about the solid lines
are apparent.
\label{fig:constraint}}
\end{center}
\end{figure}

{}Fixing the values of the non-analytic contributions to their
model-independent values, we have four remaining parameters to be
constrained for each of the regularisation schemes. All of the
regularisation schemes are constrained simultaneously to fit the
physical nucleon mass as well as the lattice QCD data. The resultant
curves are displayed in Fig.~\ref{fig:constraint}.
To the naked eye all of the curves are very much in agreement with
each other. All of them are able to give an accurate description of
the lattice data and match the physical value of $M_N$.

Now the question that we pose is: if any one of these curves were
presumed to produce an exact description of the mass of the nucleon as
a function of $m_\pi$, how well could an alternate regularisation
scheme match it? All forms have been based on the same, low-energy
effective field theory and therefore will be equivalent in the limit
$m_q\to 0$. Hence, an additional question of more direct practical
importance is: over what range can a particular regularisation scheme
match an alternate scheme? Within the radius of convergence, physical
conclusions must be independent of regularisation and renormalisation.

We show the best fit parameters for our constraint curves, $M(\mpi)$,
in Table~\ref{tab:cfit}.
\begin{table}
\begin{center}
\begin{ruledtabular}
\begin{tabular}{lccccc}
Regulator	& $a_0$    & $a_2$    & $a_4$    & $a_6$    & $\Lambda$ \\
\hline
DR		& $0.882$  & $3.82$   & $6.65$   & $-4.24$  & --        \\
BP		& $0.825$  & $4.37$   & $9.72$   & $-2.77$  & --        \\
SC		& $1.03$   & $1.12$   & $-0.292$ & --       & $0.418$   \\
MON		& $1.56$   & $0.884$  & $-0.204$ & --       & $0.496$   \\
DIP		& $1.20$   & $0.972$  & $-0.229$ & --       & $0.785$   \\
GAU		& $1.12$   & $1.01$   & $-0.247$ & --       & $0.616$   \\
\end{tabular}
\end{ruledtabular}
\caption{Bare fit parameters obtained for the constraint curves,
$M(\mpi)$. All quantities in appropriate powers of GeV. Regularisation
schemes include dimensional regularisation (DR), dimensional
regularisation maintaining the correct $\Delta\pi$ branch-point (BP),
sharp cut-off (SC), monopole (MON), dipole (DIP) and Gaussian (GAU).
\label{tab:cfit}}
\end{center}
\end{table}
It is worth recalling at this stage that the parameters listed in this
table are bare quantities and hence {\em renormalisation scheme
dependent}. If one is to rigorously compare the parameters of the
effective field theory, the self-energy contributions need to be
Taylor expanded about $\mpi=0$ in order to yield the renormalisation
for each of the coefficients in the quark mass expansion about the
chiral limit. We refer to the appendix for the details of this
procedure. A comparison of the resulting quark-mass expansion for each
of the regularisation schemes is shown in Table~\ref{tab:cren}. The
most remarkable feature of Table~\ref{tab:cren} is the very close
agreement between the values of the renormalised coefficients,
especially for the finite-range regularisation (FRR) schemes. For
example, whereas the variation in $a_0$ between all four FRR schemes
is 50\%, the variation in $c_0$ is less than half of one percent. For
$a_2$ the corresponding figure is 30\% compared with 9\% variation in
$c_2$. If one excludes the less physical sharp cut-off (SC) regulator,
the monopole, dipole and Gaussian results for $c_2$ vary by only
2\%. Finally, for $c_4$ the agreement is good for the latter three
schemes.

The comparison between $a_4$ and $c_4$ is especially important in
order to understand why the FRR schemes are so efficient. Whereas the
renormalised coefficients are consistently very large for the three
smooth FRR schemes, the bare coefficients of the residual expansion
are a factor of 100 smaller!  That is, once one incorporates the
effect of the finite size of the nucleon by smoothly suppressing pion
loops for pion masses above 0.4-0.5 GeV, the residual series expansion
has vastly improved convergence properties.

In contrast, in order to fit the nucleon mass data over an extended
range of pion mass, the dimensional regularisation schemes require
bare expansion coefficients which are much larger, 30 times larger in
the case of $a_4$. Still these coefficients are not large enough to
reach the consistent values of the smooth FRR results reported in
Table~\ref{tab:cren}. The failure of this method is amply illustrated
by the incorrect behaviour of $M_N$ as soon as one applies it even
slightly outside the fit region -- c.f. the behaviour of DR and BP for
$m_\pi^2$ between 0.8 and 1.0 GeV$^2$ shown in Fig.~\ref{fig:constraint}.
\begin{table}
\begin{center}
\begin{ruledtabular}
\begin{tabular}{lccc}
Regulator       & $c_0$    & $c_2$    & $c_4$    \\
\hline
DR              & $0.882$  & $3.82$   & $6.65$   \\
BP              & $0.885$  & $3.64$   & $8.50$   \\
SC              & $0.894$  & $3.09$   & $13.5$   \\
MON             & $0.898$  & $2.80$   & $23.6$   \\
DIP             & $0.897$  & $2.84$   & $22.0$   \\
GAU             & $0.897$  & $2.87$   & $20.7$   \\
\end{tabular}
\end{ruledtabular}
\caption{Renormalised chiral expansion parameters for the constraint
curves, $M(\mpi)$. All quantities in appropriate powers of
GeV. Regularisation schemes include dimensional regularisation (DR),
dimensional regularisation maintaining the correct $\Delta\pi$
branch-point (BP), sharp cut-off (SC), monopole (MON), dipole (DIP)
and Gaussian (GAU).
\label{tab:cren}}
\end{center}
\end{table}
%


\section{Comparison of Various Regularisation Schemes}
With these given constraint curves, $M(\mpi)$, we wish to test how
well an alternative regularisation technique can reproduce the same
curve. For any well-defined quantum
field theory, all physical results should be independent of the
regularisation and renormalisation schemes. By doing a one-to-one
comparison over different ranges of pion mass we are able to determine
the effective convergence range of each scheme.

\begin{figure}[p]
\begin{center}
{\epsfig{file=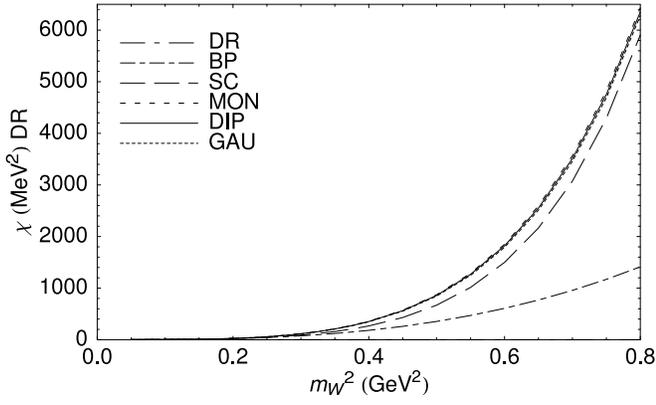, width=\smallfigsize}}
\caption{RMS area between the dimensionally regulated constraint curve
and other regularisation schemes, $\chi$, plotted as a function of the
curve fitting window $(0,m_W^2)$.
\label{fig:chiDR}}
\end{center}
\end{figure}
\begin{figure}[p]
\begin{center}
{\epsfig{file=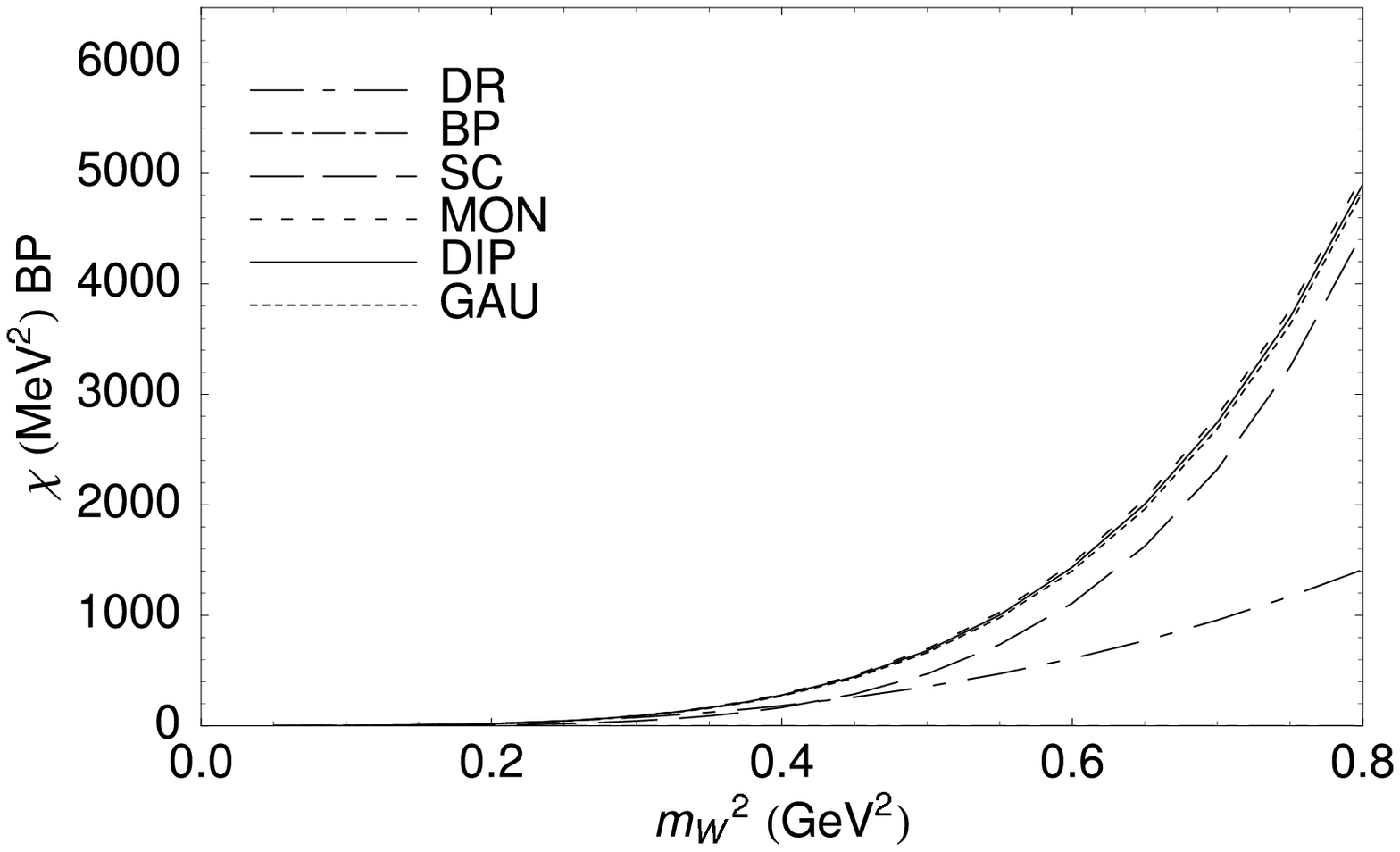, width=\smallfigsize}}
\caption{RMS area between the improved dimensionally regulated
constraint curve and other regularisation schemes, $\chi$, plotted as
a function of the curve fitting window $(0,m_W^2)$.
\label{fig:chiBP}}
\end{center}
\end{figure}
\begin{figure}[p]
\begin{center}
{\epsfig{file=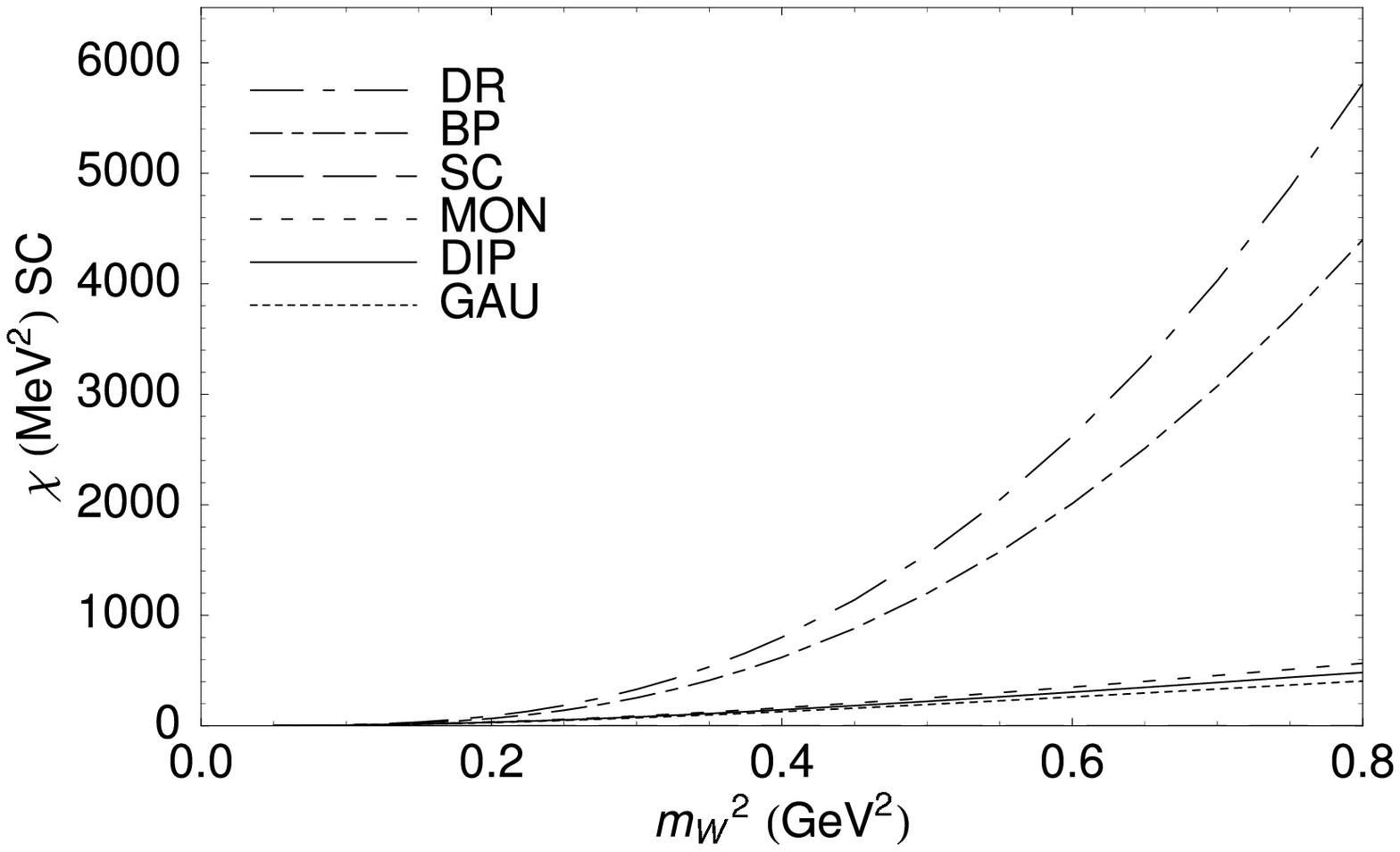, width=\smallfigsize}}
\caption{RMS area between the sharp cut-off regulated constraint curve
and other regularisation schemes, $\chi$, plotted as a function of the
curve fitting window $(0,m_W^2)$.
\label{fig:chiSC}}
\end{center}
\end{figure}
\begin{figure}[p]
\begin{center}
{\epsfig{file=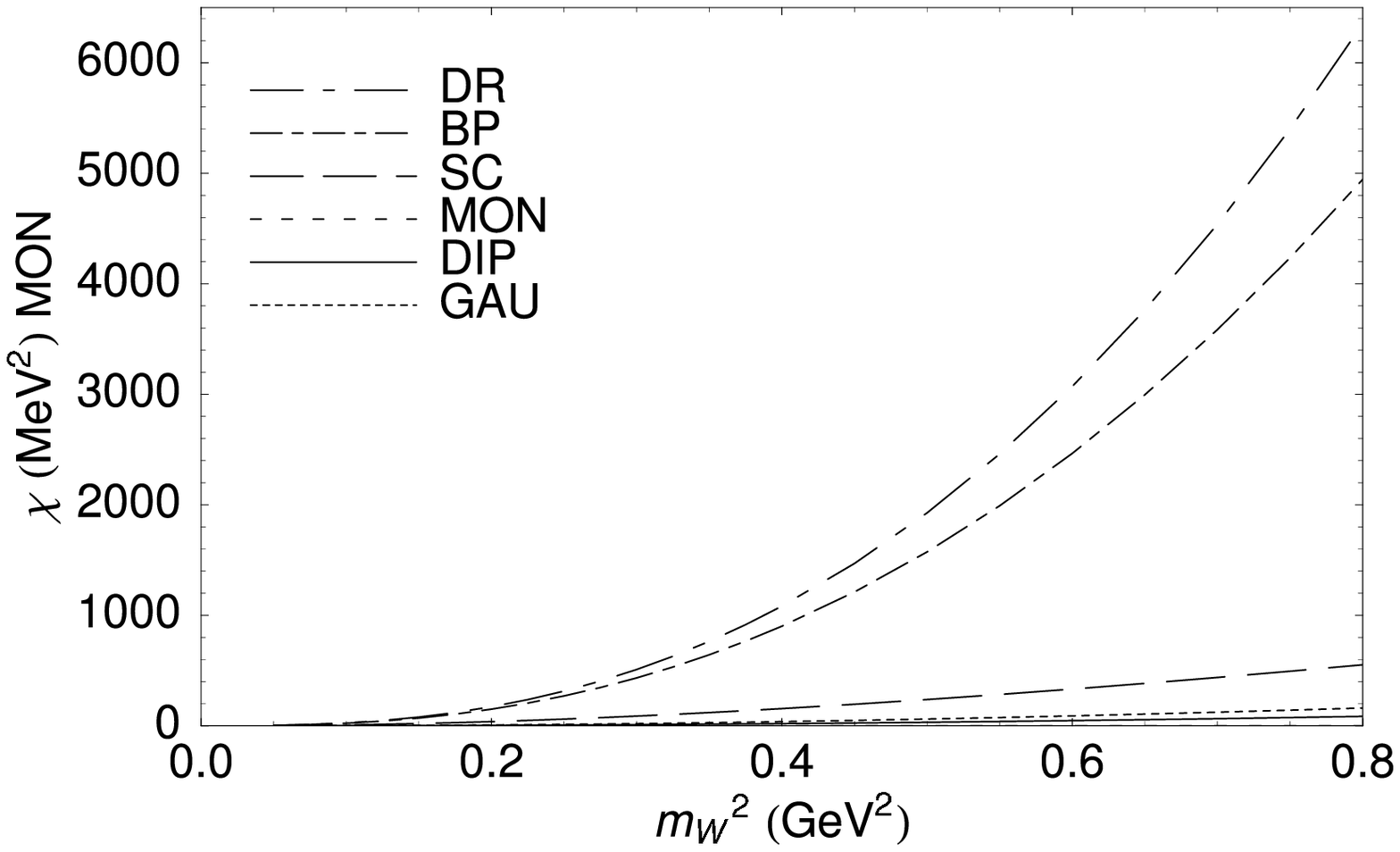, width=\smallfigsize}}
\caption{RMS area between the monopole regulated constraint curve
and other regularisation schemes, $\chi$, plotted as a function of the
curve fitting window $(0,m_W^2)$.
\label{fig:chiMON}}
\end{center}
\end{figure}
\begin{figure}[p]
\begin{center}
{\epsfig{file=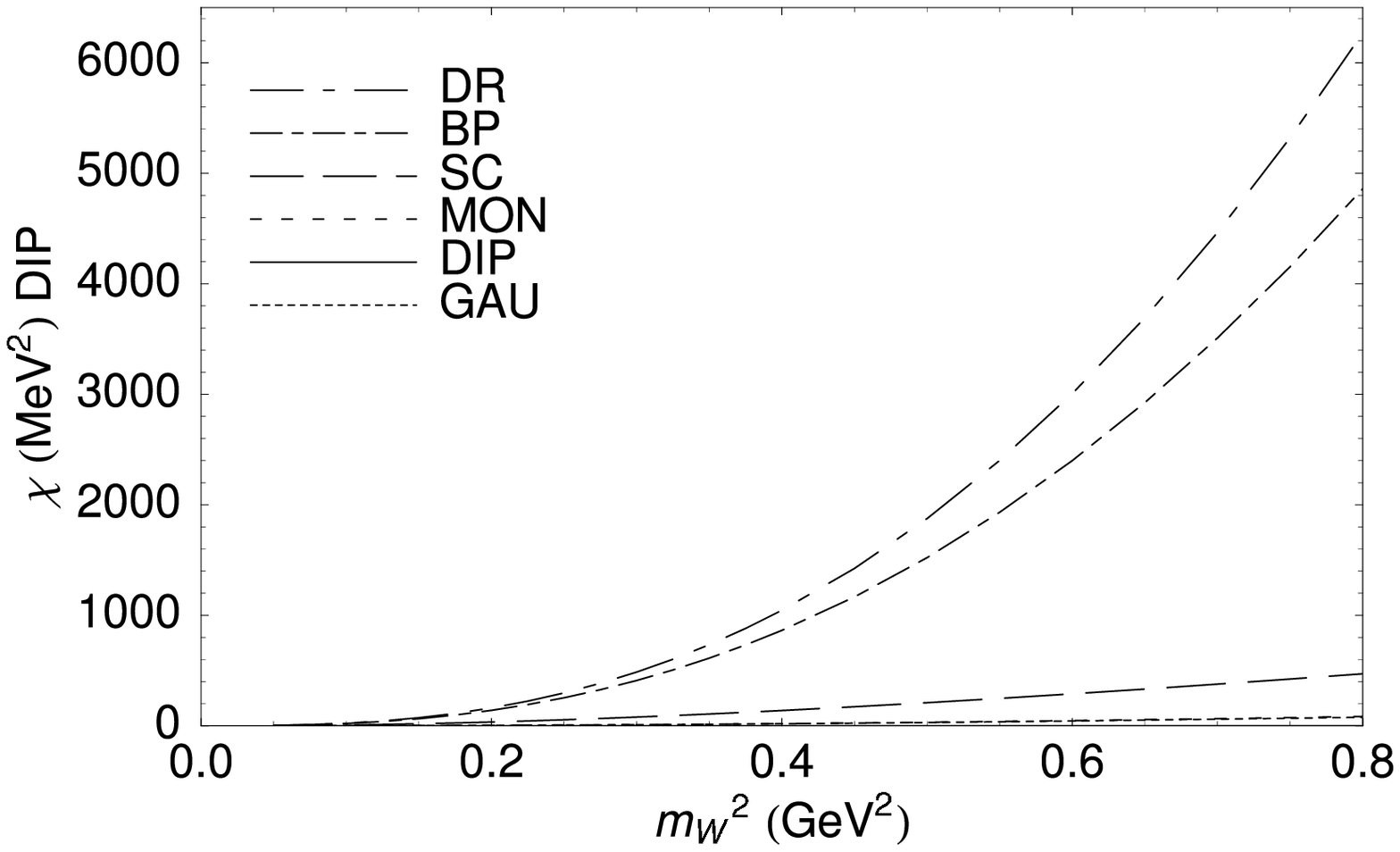, width=\smallfigsize}}
\caption{RMS area between the dipole regulated constraint curve
and other regularisation schemes, $\chi$, plotted as a function of the
curve fitting window $(0,m_W^2)$.
\label{fig:chiDIP}}
\end{center}
\end{figure}
\begin{figure}[p]
\begin{center}
{\epsfig{file=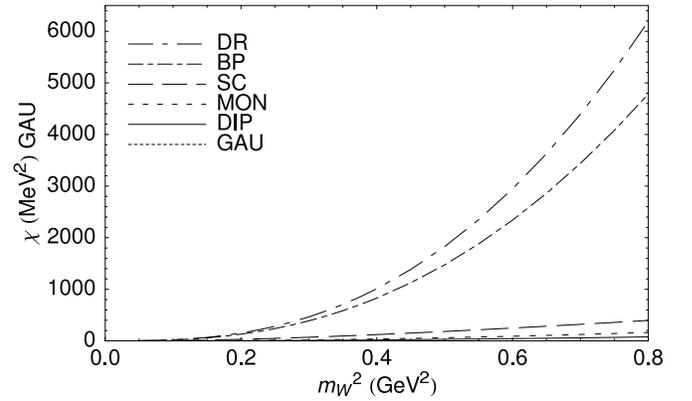, width=\smallfigsize}}
\caption{RMS area between the Gaussian regulated constraint curve
and other regularisation schemes, $\chi$, plotted as a function of the
curve fitting window $(0,m_W^2)$.
\label{fig:chiGAU}}
\end{center}
\end{figure}
With one constraint curve, $M(\mpi)$, we have five alternate curves,
$F(\mpi)$, containing free parameters, which may be adjusted to fit
this constraint. We fit each particular regularisation scheme to our
constraint curve over some window, $\mpi \in (0,m_W)$. In doing so we
define a measure, $\chi$, which describes the root-mean-square area
between the curves
\begin{equation}
\chi = \sqrt{\int_0^{m_W^2} d\mpi^2\, \left[ F(\mpi) - M(\mpi) \right]^2} \, .
\end{equation}
In practical calculations the integral over $\mpi^2$ is approximated
by the Riemann sum by dividing the region into $N$ segments of size
$h=m_W^2/N$,
\begin{equation}
\int_0^{m_W^2} d\mpi^2\, f(\mpi^2) \to h\, \sum_{i=1}^{N} f(i h) \, ,
\end{equation}
where $h$ is fixed to $0.001\gev^2$. For each upper limit, $m_W^2$,
the measure $\chi$ is minimised in the parameter space of the test
function form, $F(\mpi)$.

All regularisation prescriptions have precisely the same structure in
the limit $\mpi\to 0$. As our test window moves out to larger pion
mass the variation of $\chi$ will describe the utility of each
expansion. As an indicative scale, with $m_W^2=0.5\gev^2$ a value of
$\chi=700\mev^2$ means that the fit function is, on average, within
$1\mev$ of the constraint curve. Accurate reproduction of the
expansion coefficients also serves to test the convergence over a
given range.

The variation of $\chi$ as a function of the upper limit of the curve
matching window, $m_W$, is plotted in Figs.~\ref{fig:chiDR} to
\ref{fig:chiGAU}.  The first feature that one notices is that, if
either dimensionally-regularised scheme is used as the constraint
curve, {\it all} schemes are able to describe it very well for
$m_\pi^2$ up to 0.4 GeV$^2$.
On the other hand, we see from
Figs.~\ref{fig:chiSC} through \ref{fig:chiGAU} that neither DR nor BP
are able to reproduce the behaviour of the FRR constraint curves
beyond about 0.25 GeV$^2$. However, all of the FRR schemes are able to
reproduce the other FRR schemes over a much wider range -- in fact,
they reproduce them extremely well over the entire range of pion mass
considered (up to 0.8 GeV$^2$). Again, the coefficients in
Table~\ref{tab:cren} help us to understand this observation: it is a
simple consequence of the improved convergence properties of the
residual series expansion once any reasonable FRR scheme is
implemented.
%


\section{Chiral Coefficients}
\begin{figure}[t]
\begin{center}
{\epsfig{file=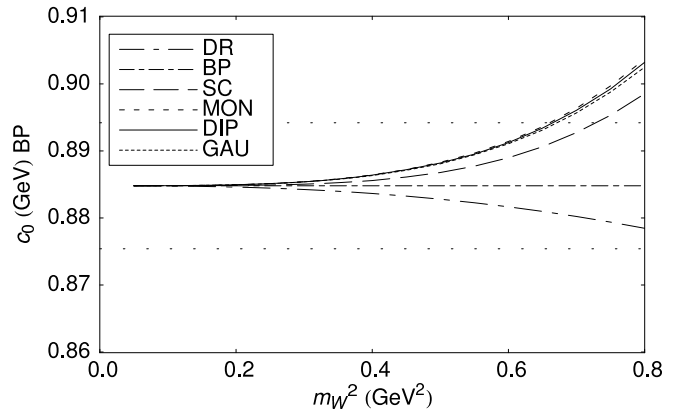, width=\myfigsize}}
\caption{Recovered $c_0$ from various regularisation schemes
constrained to the dimensionally regularised (BP) curve. $c_0$ is
plotted as a function of the curve-fitting window $(0,m_W^2)$. The
dashed horizontal line indicates the maximum deviation of $c_0$ to
within 1\% of the physical nucleon mass.
\label{fig:c0BP}}
\end{center}
\end{figure}

\begin{figure}[t]
\begin{center}
{\epsfig{file=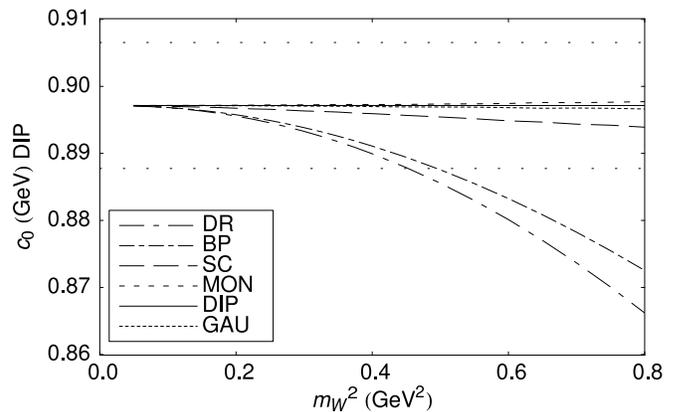, width=\myfigsize}}
\caption{Recovered $c_0$ from various regularisation schemes
constrained to the dipole regularised curve. $c_0$ is plotted as a
function of the curve-fitting window $(0,m_W^2)$. The dashed
horizontal line indicates the maximum deviation of $c_0$ to within 1\%
of the physical nucleon mass.
\label{fig:c0DIP}}
\end{center}
\end{figure}

We are concerned with the chiral expansion properties about the limit
of vanishing quark mass. It is important to test how well the
low-energy expansion parameters are reproduced as the curve-matching
window is increased. Firstly, we take the dimensionally regularised
curve as our constraint --- i.e., we assume this is the most accurate
description of the real world. We show in Fig.~\ref{fig:c0BP} the
renormalised expansion parameters, $c_0$, of the other regulators
constrained to the BP curve, as a function of the curve-fitting
window. Figure \ref{fig:c0DIP} shows a similar plot for the case where
the dipole regulator is taken as the constraint curve.

It is clear from Fig.~\ref{fig:c0BP} that, whatever scheme is used to
extract $c_0$ from the BP constraint curve, it is determined with an
accuracy better than 1\% as long as the fitting window is smaller than
0.7 GeV$^2$. We define our 1\% cut such that the contribution to the
physical nucleon mass is less than 1\% (e.g. in the case of $c_2$,
this requires that the error in the quantity $c_2 \mpi^2 / M_N < 1\%$
at the physical point).

Conversely, Fig.~\ref{fig:c0DIP} shows that the dimensional
regularisation schemes fail to yield the correct value of $c_0$ at the
1\% level once the fitting window to the dipole constraint curve
extends beyond 0.5 GeV$^2$. Yet all the FRR schemes are accurate at a
level better than half a percent, whatever window is chosen --- with
the three smooth schemes accurate to 0.1\%.

\begin{figure}[t]
\begin{center}
{\epsfig{file=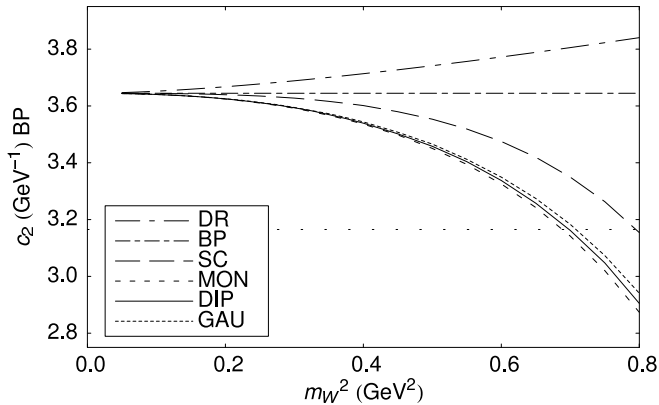, width=\myfigsize}}
\caption{Recovered $c_2$ from various regularisation schemes
constrained to the dimensionally regularised (BP) curve. $c_2$ is
plotted as a function of the curve-fitting window $(0,m_W^2)$.  The
dashed horizontal line indicates the maximum deviation $c_2$ can take
such that the error in the quantity $c_2\mpi^2$ is less than 1\% of
the physical nucleon mass at the physical pion mass.
\label{fig:c2BP}}
\end{center}
\end{figure}

\begin{figure}[t]
\begin{center}
{\epsfig{file=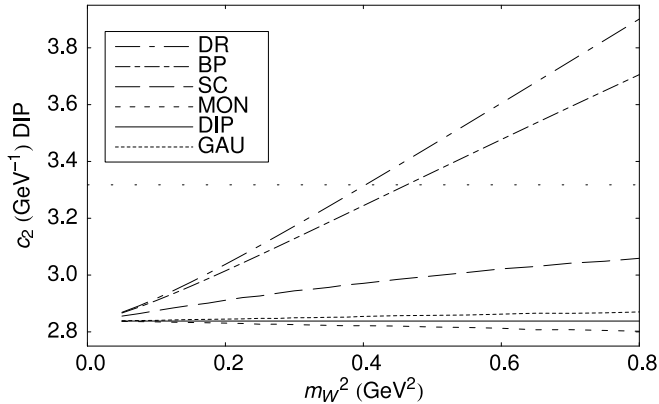, width=\myfigsize}}
\caption{Recovered $c_2$ from various regularisation schemes
constrained to the dipole regularised curve. $c_2$ is plotted as a
function of the curve-fitting window $(0,m_W^2)$. The dashed
horizontal line is as described in Fig.~\ref{fig:c2BP}.
\label{fig:c2DIP}}
\end{center}
\end{figure}

A similar story applies to the coefficient $c_2$, shown in
Figs.~\ref{fig:c2BP} and \ref{fig:c2DIP}. All methods yield the
correct value of $c_2$ based on the BP constraint curve within 1\% out
to a fit window of 0.7 GeV$^2$. The breakdown at this scale should
come as little surprise, since the highest lattice data point used to
constrain the BP curve lies at around 0.7 GeV$^2$.

When matching to the dipole constraint (Fig.~\ref{fig:c2DIP}), the
dimensionally regularised cases break down much earlier, at around 0.4
GeV$^2$. All three smooth FRR schemes reproduce this coefficient over
the entire range to similar accuracy as found for $c_0$.

As we commented in Section~\ref{sec:constrain}, consistent results are
found for the renormalised coefficients $c_4$ over the entire fit
range $0\le \mpi^2 \le 0.8\gev^2$, provided a smooth finite-range
regulator is used to regulate the effective field theory. $c_4$ is
compromised in the dimensionally-regulated schemes as it attempts to
simulate higher-order terms absent in the truncated expansion.

It is clear from these exercises that one cannot hope to use a series
expansion based upon dimensional regularisation to analyse data over a
window wider than 0.4 GeV$^2$. That means one would need to have
sufficient, accurate lattice QCD data in this mass region to fix four
fitting parameters before one could hope to trust the chiral
coefficients so obtained. Within this region, it would also be
necessary to have data very near the physical pion mass.  This is
certainly beyond the likely computational capacity of current
collaborations in the next 5-10 years.  On the other hand, it is
apparent that by using the improved convergence properties of the FRR
schemes one can use lattice data in the region up to 0.8 GeV$^2$ or
greater. This is a regime where we already have impressive data from
CP-PACS \cite{AliKhan:2001tx} and MILC \cite{Bernard:2001av}. In
particular, the FRR approach offers the ability to extract reliable
fits where the low-mass region is excluded from the available
data. The consideration of this practical application to the
extrapolation of lattice data will be discussed further below.

Indeed, these results suggest that given data from the next generation
of (10 Teraflops) lattice QCD machines currently under construction,
the FRR schemes should allow us to extract $c_0$ and $c_2$,
independent of any model, at the 1\% level and $c_4$ at the level of a
few percent.


\section{Discussion and Conclusions}
We have observed that all the regularisation schemes produce
consistent results for $c_0$ and $c_2$, with an accuracy sufficient to
predict the nucleon mass at the 1\% level, provided the expansion is
limited to pion masses within the range $0 \le m_\pi^2 \le 0.4\ {\rm
GeV}^2$.  The curves are very similar with an RMS area between pairs
of curves less than 0.001 GeV${}^2$.

However, it is the dimensionally-regulated curves that are unable to
accurately reproduce the results of the finite-range regulated curves
beyond the range $0 \le m_\pi^2 \le 0.4\ {\rm GeV}^2$.  As
contributions from higher order terms not included in the truncated
dimensionally-regulated expansion become important, the coefficients
of the existing terms move away from their correct values in order to
compensate for missing terms.

On the contrary, the finite-range regulated curves are able to
reproduce the predictions of dimensional regularisation at the 1\%
level over a wider range approaching $0 \le m_\pi^2 \le 0.7\ {\rm
GeV}^2$.  This enhanced range of the finite-range regulated curves
is largely a result of the flexibility afforded by the presence of a
separate regulation scale, $\Lambda$, which may be adjusted while
preserving the excellent convergence properties of the finite-range
regulated effective field theory. Hence, one can conclude that the
applicable range of {\it finite-range regulated} effective field
theory in which model independent conclusions may be drawn is $0 \le
m_\pi^2 \le 0.7\ {\rm GeV}^2$. This range is sufficient to solve the
problem of formulating $SU(3)$-flavor chiral perturbation theory where
the strange quark gives rise to a pseudoscalar kaon mass of 0.5 GeV.

However, it is the predictions of dimensional regularisation that
prevent this range in pion mass from being larger. Indeed the
predictions of each of the smooth finite-range regulated formulations
agree at an extraordinarily precise level over the entire range of
pion mass considered, $0 \le m_\pi^2 \le 0.8\ {\rm GeV}^2$. This
suggests that the more restricted regimes discussed above are
associated with a fault in the choice of dimensional regularisation as
a regulator for effective field theory in this context.

It is easy to see why this may be. In dimensionally-regularised
$\chi$PT, the mass of the pseudoscalar governs the
scale of physics contributing to the loop integrals of $\chi$PT. As
the pseudoscalar mass increases, the dimensionally-regularised results
become dominated by short distance physics; precisely the regime where
effective field theory fails because of the internal structure of the
hadrons represented by the effective fields. 
Ignoring this physics by using point-like couplings leads to a badly divergent 
series expansion at higher order which, in turn, makes the method 
unsuitable for the lattice extrapolation problem in the foreseeable future.

Fortunately this incorrect short-distance physics, included in
dimensionally-regulated loop integrals, can be removed by a more
appropriate choice of regulator and a corresponding shift of the
parameters of the chiral Lagrangian. In the region where these
expansions both converge their predictions must agree as long as the
series are not truncated too soon. In a truncated expansion, these
incorrect contributions are not removed. In dimensionally-regularised
$\chi$PT they become large for increasing $m_\pi$,
introducing a significant model dependence in the truncated chiral
expansion and giving rise to the catastrophic failure already apparent
in Fig.~1 for $m_\pi^2 > 0.8\ {\rm GeV}^2$. Convergence of the
dimensionally-regulated expansion is slow as large errors, associated
with short-distance physics in loop integrals (not suppressed in
dimensionally-regulated $\chi$PT), must be removed by equally large
analytic terms.

The enhanced pion-mass range of the finite-range regulated curves is a
consequence of the effective resummation of the chiral expansion that
arises via the regulator. For example, a Taylor series expansion of
the dipole-regulated result for the $N \to N \pi$ self energy of
Eq.~(\ref{eq:IpiDIP}) in the Appendix reveals not only the standard
LNA $m_\pi^3$ contribution, but also new non-analytic contributions
proportional to $m_\pi^5$ and higher powers. As shown by Birse and
McGovern, the coefficient of the $m_\pi^5$ term is related to the
Goldberger-Treiman (GT) discrepancy~\cite{McGovern:1998tm}\footnote{In
dimensionally regulated \chpt\ it arises only at the two-loop level,
whereas in FRR it is already present at one loop due to resummation of
the chiral expansion via the finite-ranged regulator. Furthermore, it
is well known that a form factor at the $\pi NN$ vertex with mass
parameters in the range shown in Table~\ref{tab:cfit} leads to a
correction to the GT relation of the same size as the observed
discrepancy.}. The coefficients of these new terms involve the dipole
regulator parameter $\Lambda$ which can be optimally constrained in
the fitting procedure to reproduce the correct non-analytic structure
displayed in the data. (Indeed, the sign and magnitude of the
coefficient of $m_\pi^5$ that we find agrees with the value found by
Birse and McGovern -- although the latter has relatively large
errors.)  Such terms can only arise at higher order in the
dimensionally-regulated expansion as there is no {\it effective}
re-summation of the series.

It is interesting to note that the appearance of higher-order
non-analytic structure, such as the $m_\pi^5$ term discussed above,
does not occur for the special case of the sharp cut-off regulator.
An expansion of the arctangent in Eq.~(\ref{eq:IpiSC}) provides a
constant followed by odd powers of $m_\pi$ such that when multiplied
by $m_\pi^3$ only analytic terms appear.  This unfortunate feature of
the sharp-cutoff regulator explains why it lies between
dimensional-regularisation and the smooth finite-range regulated
results.  While analytic terms can be effectively re-summed, the
sharp-cutoff regulator suffers the same restrictions as dimensional
regularisation, in that higher-order non-analytic behavior is not
effectively re-summed.

The precise agreement between the smooth finite-range regulated
results over the entire pion-mass range considered, $0 \le m_\pi^2 \le
0.8\ {\rm GeV}^2$, confirms that the shape of the regulator is
irrelevant provided that the regulator parameter is optimised by the
fit to the data.  An optimal regulator (perhaps guided by
phenomenology) effectively re-sums the chiral expansion, encapsulating
the essential physics in the first few terms of the expansion.  The
approach is systematically improved by simply going to higher order in
the chiral expansion, introducing additional analytic terms as
appropriate to maintain model independence.

It is often argued that dimensionally regularisation works well in the
meson sector. However, we know of no reason that the findings
discussed here for the baryon sector should not apply equally well
there. Improved convergence of finite-range regulated \chpt\ for meson
properties will provide access to a wide range of quark masses and
precision determinations of the low-energy constants.


As discussed in the introduction it of great interest how to apply the
features of effective field theory to the extrapolation of lattice QCD
results. If one wishes to extract the low-energy constants from
currently accessible lattice simulations, it is necessary that any
such prescription is not sensitive to the choice of regularisation or
renormalisation. Here we describe how the FRR schemes achieve this
result.

\begin{figure}[t]
\begin{center}
{\epsfig{file=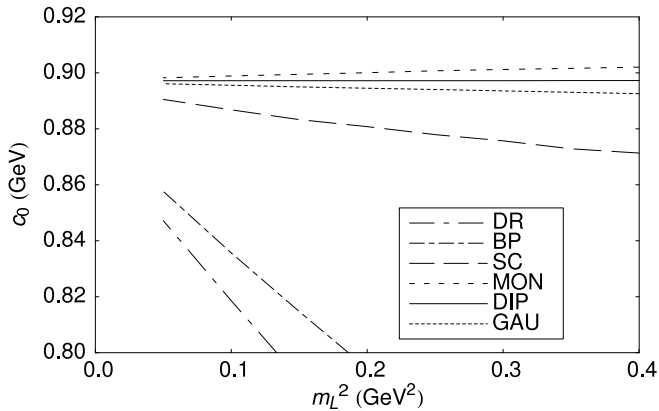, width=\myfigsize}}
\caption{Extrapolated $c_0$ as function of lowest {\em data}
point. Data points are spaced $0.05\gev^2$ apart and the maximum point
is $0.8\gev^2$.
\label{fig:Extc0}}
\end{center}
\end{figure}

\begin{figure}[t]
\begin{center}
{\epsfig{file=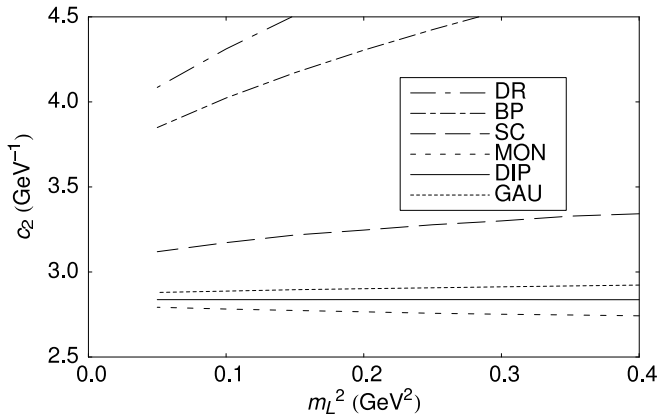, width=\myfigsize}}
\caption{Extrapolated $c_2$ as function of lowest {\em data}
point. Data points are spaced $0.05\gev^2$ apart and the maximum point
is $0.8\gev^2$.
\label{fig:Extc2}}
\end{center}
\end{figure}

With current lattice data typically restricted to $\mpi^2>0.3\gev^2$
one must have a sufficiently wide window in order to accurately
determine four fit parameters of the chiral expansion. Thus for the
test of model-dependence in extrapolated results we presume perfect
data up to our maximum considered range, $\mpi^2=0.8\gev^2$. We fix
the pseudo-data by the dipole regularised curve, the dipole being
chosen as one of the three regulators which give best agreement over
the widest range. Data points are chosen to be equally spaced by
$0.05\gev^2$, from $0.8\gev^2$ down to some minimum available point at
$m_L^2$. Figures~\ref{fig:Extc0} and \ref{fig:Extc2} show the
reconstructed low-energy constants from the fitting window
($m_L^2$,$0.8\gev^2$) with $m_L^2$ in the range $0.4 \ge m_L^2 \ge
0.05\gev^2$.

It is clear that the dimensionally regularised extrapolation schemes
perform poorly. Even with {\em perfect} lattice QCD results down to
$0.1\gev^2$, $c_0$ is likely to be in error by 7\% and $c_2$ by around
40\%. On the other hand, one finds excellent agreement in the chiral
parameters, $c_0$ and $c_2$, obtained from the finite-ranged
regulators, even with the lowest data point situated at
$0.4\gev^2$. This indicates that, using the technique of finite-range
regularisation, one should be able to extract chiral coefficients at
the 1\% level from the lattice QCD data which will be obtained in the
next few years.


\section*{Acknowledgments}
We would like to thank C.~R.~Allton, M.~Birse, J.~McGovern and
S.~V.~Wright for helpful conversations. We also thank W.~Melnitchouk
for careful reading of the manuscript.
This work was supported by the Australian Research Council and the
University of Adelaide.


\appendix

\section{Renormalisation}
Here we summarise the renormalisation prescription for various
regularisation schemes. There are two types of diagrams which need to
be considered in this analysis. The first of these is the $NN\pi$-type
diagram, while the second corresponds to the case of a non-degenerate
intermediate state, like the $N\Delta\pi$.

Within the heavy-baryon, non-relativistic approximation the integral
describing the $N\to N\pi$ diagram is written as
\begin{equation}
I_\pi = \frac{2}{\pi} \int_0^\infty dk\, \frac{k^4\, u^2(k)}{k^2+m^2}\, ,
\end{equation}
where the factor $u(k)$ is the $N\pi$ vertex regulator. The integral
has been defined in such a way that the coefficient of the $\mpi^3$
contribution is normalised to unity. For the off-diagonal
contribution the loop-integral can be expressed as
\begin{equation}
I_{\pi\Delta} = \frac{2}{\pi} \int_0^\infty dk\, 
\frac{k^4\, u^2(k)}{\sqrt{k^2+m^2}(\Delta+\sqrt{k^2+m^2})}\, ,
\end{equation}
with $\Delta$ the $\Delta$--$N$ mass-splitting, defined as positive for a heavier
intermediate state. The normalisation is chosen to agree with the
previous equation in the limit $\Delta\to 0$. The various functional
forms for the finite-ranged regulators are listed in the text.

Independent of the regulator, the Taylor expansion of the integral,
$I_\pi$, behaves as
\be
I_\pi(m\sim 0) = b_0 + b_2 m^2 + m^3 + b_4 m^4 + \dots
\ee
With the intermediate state being non-degenerate, the $m^3$
contribution becomes a logarithm and the expansion is given by
\bea
I_{\pi\Delta}(m\sim 0) = & b_{0\Delta} + b_{2\Delta} m^2 + b_{4\Delta} m^4 \non\\
\noalign{\smallskip}
      & -\frac{3}{4\,\pi\,\Delta} m^4 \log m + \ldots
\eea
With respect to the chiral expansion, Eq.~(\ref{eq:eftexp}), the
self-energy contributions giving rise to the LNA and NLNA are given by
\bea
\sigma_{{\rm N} \pi}(m_\pi, \Lambda) &=& C_N\, I_\pi(\mpi,\Lambda)\, , 
\non\\
\sigma_{\Delta \pi}(m_\pi, \Lambda)  &=& C_\Delta\, I_{\pi\Delta}(\mpi,\Lambda)\, , 
\eea
where for simplicity we have defined $C_N=c_{LNA}$ and
$C_\Delta=-c_{NLNA}4\pi\Delta/3$. The renormalisation to obtain the
low-energy constants of the Taylor expansion, Eq.~(\ref{eq:drexp}), is
then simply given by
\bea
c_0 & = & a_0 + C_N\, b_0 + C_\Delta\, b_{0\Delta} \, ,\non \\
c_2 & = & a_2 + C_N\, b_2 + C_\Delta\, b_{2\Delta} \, ,\non \\
c_4 & = & a_4 + C_N\, b_4 + C_\Delta\, b_{4\Delta} \, .
\eea
It is these renormalised coefficients, $c_i$, which have physical
significance and it is these that should be compared with
phenomenological studies of chiral expansions.

In Table~\ref{tab:bi} we summarise the $b_i$ expansion
coefficients. We were unable to obtain neat analytic expressions for
the Gaussian regulator and hence all relevant calculations have been
performed numerically. For both DR and BP schemes the $b_i$ of the
$I_\pi$ diagram can trivially be treated as zero.
\begin{table}
\begin{center}
\begin{tabular}{lrrr}
\hline\hline
Regulator\hspace{2mm}  & \hspace{10mm}$b_0$ & \hspace{10mm}$b_2$ & \hspace{10mm}$b_4$    \\
\hline
\noalign{\smallskip}
Sharp       & $\frac{2 \Lambda^3}{3 \pi}$
            & $- \frac{2 \Lambda }{\pi}$
            & $- \frac{2}{\pi \Lambda}$             \\
\noalign{\smallskip}
Monopole    & $\frac{\Lambda^3}{2}$
            & $- \frac{\Lambda }{2}$
            & $- \frac{3}{2 \Lambda}$               \\
\noalign{\smallskip}
Dipole      & $\frac{\Lambda^3}{16}$
            & $- \frac{5 \Lambda }{16}$
            & $- \frac{35}{16 \Lambda}$             \\
\noalign{\smallskip}
\hline\hline
\end{tabular}
\caption{Expansion coefficients for the finite regulators.
\label{tab:bi}}
\end{center}
\end{table}
The expansion terms, $b_{i\,\Delta}$ for the $I_{\pi\Delta}$ diagram
are more complicated and are summarised in the following.

In the DR case, the $b_{i\,\Delta}$ can again be treated as zero. From
the expansion of Eq.~(\ref{eq:log}), for the BP case one finds that the appropriate
coefficients are given by (with renormalisation scale $1.0\gev$)
\bea
b_{0\Delta}^{\rm BP} &=& \frac{\Delta^3}{\pi}\,\log(4\Delta^2) \, , \non \\
b_{2\Delta}^{\rm BP} &=& -\frac{\Delta}{2\,\pi}\,\left[1+3\log(4\Delta^2) \right] \, , \non \\
b_{4\Delta}^{\rm BP} &=& \frac{3}{16\,\pi\,\Delta}\,\left[3+4\log2 + 2 \log(\Delta^2)\right] \, .
\eea
\begin{widetext}
For the sharp cut-off, expressions are given by
\bea
b_{0\Delta}^{\rm SC} &=& \frac{1}{3\pi} \left[ 6\Delta^2\Lambda - 3\Delta\Lambda^2 + 2\Lambda^3 
                 + 6\Delta^3 \log{\frac{\Delta}{\Delta+\Lambda}} \right]\, , \non \\
b_{2\Delta}^{\rm SC} &=& \frac{1}{\pi (\Delta + \Lambda)} \left[ -\Lambda(3\Delta+2\Lambda) 
                -3\Delta(\Delta+\Lambda) \log{\frac{\Delta}{\Delta+\Lambda}} \right]\, , \non \\
b_{4\Delta}^{\rm SC} &=& \frac{-1}{16\pi\Delta(\Delta + \Lambda)^2} 
               \left[ 7\,\Delta^2 + 2\,\Delta\,\Lambda - 9\,\Lambda^2 
               + 12\,( \Delta + \Lambda )^2\,
               \log{\frac{\Delta + \Lambda}{2\,\Delta\,\Lambda}} \right]\, .
\eea
For the monopole we obtain
\bea
b_{0\Delta}^{\rm MON} &=& \frac{\Lambda^4}{2\,\pi(\Delta^2+\Lambda^2)^2}
        \left[ -2\,\Delta^3 - 2\,\Delta\,\Lambda^2 + 3\,\Delta^2\,\Lambda\,\pi 
               + \Lambda^3\,\pi
               + 2\,\Delta^3\,\log\left(\frac{\Delta^2}{\Lambda^2}\right) \right]\, , \non \\
b_{2\Delta}^{\rm MON} &=& -\frac{\Lambda^2}{2\,\pi(\Delta^2+\Lambda^2)^3}
        \Bigg[ \Delta^5 + 6\,\Delta^3\,\Lambda^2 + 5\,\Delta\,\Lambda^4 
               - 3\,\Delta^2\,\Lambda^3\,\pi + \Lambda^5\,\pi \non \\
& & \hspace{7mm} - 2\,\Delta\,\Lambda^2\,\left( \Delta^2 - 3\,\Lambda^2 \right)
                   \,\log \left(\frac{\Delta}{\Lambda}\right) \Bigg]\, , \non \\
b_{4\Delta}^{\rm MON} &=& -\frac{1}{16\,\pi\,\Delta(\Delta^2+\Lambda^2)^4}
        \Bigg[ 13\,\Delta^8 + 54\,\Delta^6\,\Lambda^2 + 108\,\Delta^4\,\Lambda^4 
               + 58\,\Delta^2\,\Lambda^6 - 9\,\Lambda^8 - 24\,\Delta^3\,\Lambda^5\,\pi \non \\
& & \hspace{7mm} + 24\,\Delta\,\Lambda^7\,\pi 
               - 6\,\Lambda^4\,\left( \Delta^4 - 6\,\Delta^2\,\Lambda^2 + \Lambda^4 \right) \,
                 \log \left(4\Delta^2\right)  \non \\
& & \hspace{7mm} - 6\,\left( \Delta^8 + 4\,\Delta^6\,\Lambda^2 + 5\,\Delta^4\,\Lambda^4
               + 10\,\Delta^2\,\Lambda^6 \right) \,\log\left(4\Lambda^2\right) \Bigg]\, .
\eea
For the dipole we obtain
\bea
b_{0\Delta}^{\rm DIP} & = & \frac{\Lambda^4}{48\pi (\Delta^2 + \Lambda^2)^4}
                \Bigg[ 8\,\Delta^7 + 48\,\Delta^5\,\Lambda^2 + 24\,\Delta^3\,\Lambda^4 
                      - 16\,\Delta\,\Lambda^6 - 3\,\Delta^6\,\Lambda\,\pi               \non\\
& & \hspace{7mm} - 27\,\Delta^4\,\Lambda^3\,\pi  + 
                  27\,\Delta^2\,\Lambda^5\,\pi  + 3\,\Lambda^7\,\pi  + 
                  96\,\Delta^3\,\Lambda^4\,\log{\frac{\Delta}{\Lambda}} \Bigg]\, , \non \\
b_{2\Delta}^{\rm DIP} & = & -\frac{\Lambda^2}{48\pi (\Delta^2 + \Lambda^2)^5}
          \Bigg[ 8\,\Delta^9 + 36\,\Delta^7\,\Lambda^2 + 36\,\Delta^5\,\Lambda^4 
                  + 188\,\Delta^3\,\Lambda^6 + 180\,\Delta\,\Lambda^8 
                  + 3\,\Delta^6\,\Lambda^3\,\pi \non\\
& & \hspace{7mm} + 45\,\Delta^4\,\Lambda^5\,\pi  
		    - 135\,\Delta^2\,\Lambda^7\,\pi  + 15\,\Lambda^9\,\pi
                    - 48\,\left( 5\,\Delta^3\,\Lambda^6 - 3\,\Delta\,\Lambda^8 \right) 
                    \, \log{\frac{\Delta}{\Lambda}} \Bigg]\, , \non\\
a_{4\Delta}^{\rm DIP} & = &
   \frac{-1}{48\,\Delta\,{\left( \Delta^2 + \Lambda^2 \right) }^6\,\pi }
   \Bigg[ 54\,\Delta^{12} + 326\,\Delta^{10}\,\Lambda^2 + 819\,\Delta^8\,\Lambda^4 + 
           1062\,\Delta^6\,\Lambda^6 + 1154\,\Delta^4\,\Lambda^8 \non\\
& & \hspace{7mm} + 612\,\Delta^2\,\Lambda^{10} - 
      27\,\Lambda^{12} + 3\,\Delta^7\,\Lambda^5\,\pi  + 63\,\Delta^5\,\Lambda^7\,\pi
      - 315\,\Delta^3\,\Lambda^9\,\pi  + 105\,\Delta\,\Lambda^{11}\,\pi  \non\\
& & \hspace{7mm} - 6\,\Lambda^8\,\bigg( 35\,\Delta^4 - 42\,\Delta^2\,\Lambda^2 + 3\,\Lambda^4 \bigg) \,
       \log\left(4\Delta^2\right) - 6\,\Delta^2\,\bigg( 3\,\Delta^{10} + 18\,\Delta^8\,\Lambda^2 \non\\
& & \hspace{7mm} + 45\,\Delta^6\,\Lambda^4 + 60\,\Delta^4\,\Lambda^6 
      + 10\,\Delta^2\,\Lambda^8 + 60\,\Lambda^{10} \bigg) \,
       \log \left(4\Lambda^2\right) \Bigg]\, .
\eea

Finally, we give the full expressions for the finite-range regulated
integrals. For the simple diagram, $I_\pi$, we obtain
\bea
I_\pi^{\rm SC}  & = & \frac{2 \Lambda^3}{3 \pi} 
  - \frac{2 \Lambda }{\pi} m^2 + 
  \frac{2}{\pi} m^3\,\arctan \left(\frac{\Lambda}{m}\right)\, , \label{eq:IpiSC}  \\
I_\pi^{\rm MON} & = & \frac{\Lambda^4 ( 2 m + \Lambda ) }
                         {2 ( m + \Lambda )^2 }\, ,    \label{eq:IpiMON}          \\
I_\pi^{\rm DIP} & = & \frac{ \Lambda^5 ( m^2 + 4 m \Lambda + \Lambda^2 ) }
                         { 16 ( m + \Lambda )^4 }\, . \label{eq:IpiDIP}
\eea
For the more complicated integral, $I_{\pi\Delta}$, we obtain for the
sharp cut-off
\bea
I_{\pi\Delta}^{\rm SC} & = &
\frac{1}{3\pi}\Bigg\{
6 \Delta^2 \Lambda + 2 \Lambda^3 - 6 \Lambda m^2 - 3 \Delta \Lambda {\sqrt{\Lambda^2 + m^2}}
    + 3 \Delta \left( -2 \Delta^2 + 3 m^2 \right) \log (\frac{\Lambda + {\sqrt{\Lambda^2 + m^2}}}{m}) \non\\
& & \hspace{7mm} - 6 {\left( \Delta^2 - m^2 \right) }^{\frac{3}{2}} 
\Bigg[ \log\left(\frac{-\Delta - m + {\sqrt{\Delta^2 - m^2}}}
          {\Delta + m + {\sqrt{\Delta^2 - m^2}}}\right) \non\\
& & \hspace{7mm} +  \log\left(\frac{\Delta + \Lambda + {\sqrt{\Delta^2 - m^2}} + 
           {\sqrt{\Lambda^2 + m^2}}}{-\Delta - \Lambda + 
           {\sqrt{\Delta^2 - m^2}} - {\sqrt{\Lambda^2 + m^2}}}\right)
       \Bigg]\Bigg\} \, .
\eea
For the monopole regulated integral we obtain
\bea
I_{\pi\Delta}^{\rm MON} & = &
    \frac{\Lambda^4}{2 \pi (\Lambda^2 + \Delta^2 - m^2)^2} \non\\
& & \Bigg[ \frac{\Lambda (3 \pi m^4 + m^2 (-3\pi\Delta^2 + 2\Delta\Lambda -4\pi\Lambda^2) 
                + \Lambda (-2\Delta^3 +3\pi\Delta^2\Lambda -2\Delta\Lambda^2 +\pi\Lambda^3) )}
                {\Lambda^2 - m^2} \non\\
& &       - 2 (\Delta^2 - m^2)^{3/2} 
              \log\left(\frac{\Delta-\sqrt{\Delta^2-m^2}}{\Delta+\sqrt{\Delta^2-m^2}}\right) \non\\
& &       + \frac{\Delta\Lambda (3m^4 +2\Delta^2\Lambda^2 - 3m^2 (\Lambda^2 + \Delta^2) )}
                  {(\Lambda^2 - m^2)^{3/2}}
            \log\left(\frac{\Lambda-\sqrt{\Lambda^2-m^2}}{\Lambda+\sqrt{\Lambda^2-m^2}}\right) 
          \Bigg]\, .
\eea
For the dipole regulated integral we obtain
\bea
I_{\pi\Delta}^{\rm DIP} & = & \frac{\Lambda^5}{48\,
      {\left( \Lambda^2 - m^2 \right) }^{\frac{7}{2}}\,
      {\left( \Delta^2 + \Lambda^2 - m^2 \right) }^4\,\pi }
    \Bigg\{ {\sqrt{\Lambda^2 - m^2}}  \non\\
& &    \Bigg[ -16\,\Delta\,\Lambda^{11} + 
         4\,\Delta\,\Lambda^9\,\left( 6\,\Delta^2 - 11\,m^2 \right)  + 
         6\,\Delta\,\Lambda\,m^4\,{\left( -\Delta^2 + m^2 \right) }^3  \non\\
& &      + 2\,\Delta\,\Lambda^3\,m^2\,{\left( -\Delta^2 + m^2 \right) }^2
         \left( -16\,\Delta^2 + 37\,m^2 \right)  \non\\
& &      + 6\,\Delta\,\Lambda^7\,\left( 8\,\Delta^4 - 44\,\Delta^2\,m^2 + 37\,m^4 \right) \non\\
& &      - 2\,\Delta\,\Lambda^5\,\left( -\Delta + m \right) \,
          \left( \Delta + m \right) \,
          \left( 4\,\Delta^4 - 98\,\Delta^2\,m^2 + 121\,m^4 \right)          \non\\
& &      + 3\,\Lambda^{12}\,\pi  - 3\,m^6\,{\left( -\Delta^2 + m^2 \right) }^3\,
          \pi  - 9\,\Lambda^{10}\,\left( -3\,\Delta^2 + 4\,m^2 \right) \,\pi  \non\\
& &      + 9\,\Lambda^2\,m^4\,{\left( -\Delta^2 + m^2 \right) }^2\,
          \left( -\Delta^2 + 4\,m^2 \right) \,\pi  \non\\
& &      - 9\,\Lambda^4\,m^2\,\left( -\Delta + m \right) \,
          \left( \Delta + m \right) \,
          \left( \Delta^4 - 11\,\Delta^2\,m^2 + 7\,m^4 \right) \,\pi  \non\\
& &      + 9\,\Lambda^8\,\left( -3\,\Delta^4 - 3\,\Delta^2\,m^2 + 7\,m^4 \right) \,
          \pi  \non\\
& &      - 3\,\Delta^2\,\Lambda^6\,
          \left( \Delta^4 - 30\,\Delta^2\,m^2 + 30\,m^4 \right) \,\pi  \Bigg] \non\\
& &    - 48\,\Lambda^3\,{\left( \Delta^2 - m^2 \right) }^{\frac{3}{2}}\,
         {\left( \Lambda^2 - m^2 \right) }^{\frac{7}{2}}\,
         \log \left(\frac{\Delta - {\sqrt{\Delta^2 - m^2}}}
              {\Delta + {\sqrt{\Delta^2 - m^2}}}\right) \non\\
& &    - 3\,\Delta\,\Bigg[ m^6\,{\left( -\Delta^2 + m^2 \right) }^3 + 
         8\,\Lambda^{10}\,\left( -2\,\Delta^2 + 3\,m^2 \right) \non\\
& &      - 3\,\Lambda^2\,m^4\,{\left( -\Delta^2 + m^2 \right) }^2\,
          \left( -2\,\Delta^2 + 5\,m^2 \right)  \non\\
& &      - 4\,\Lambda^8\,m^2\,\left( -14\,\Delta^2 + 15\,m^2 \right)  + 
         \Lambda^6\,m^4\,\left( -34\,\Delta^2 + 35\,m^2 \right)  \non\\
& &      + 3\,\Lambda^4\,m^4\,\left( 8\,\Delta^4 - 13\,\Delta^2\,m^2 + 5\,m^4 \right) 
         \Bigg] \,\log \left(\frac{\Lambda - {\sqrt{\Lambda^2 - m^2}}}
         {\Lambda + {\sqrt{\Lambda^2 - m^2}}}\right) \Bigg\}\, .
\eea
\end{widetext}



\begin{thebibliography}{40}


\bibitem{Detmold:2001hq}
W. Detmold et~al.,
\newblock Pramana 57 (2001) 251, nucl-th/0104043.

\bibitem{Leinweber:1993hj}
D.B. Leinweber and T.D. Cohen,
\newblock Phys. Rev. D47 (1993) 2147, hep-lat/9211058.

\bibitem{Leinweber:1994yw}
D.B. Leinweber and T.D. Cohen,
\newblock Phys. Rev. D49 (1994) 3512, hep-ph/9307261.

\bibitem{ELECTRIC}
D.B. Leinweber, D.H. Lu and A.W. Thomas,
\newblock Phys. Rev. D60 (1999) 034014, hep-lat/9810005;
E.J. Hackett-Jones, D.B. Leinweber and A.W. Thomas,
\newblock Phys. Lett. B489 (2000) 143, hep-lat/0004006;
E.J. Hackett-Jones, D.B. Leinweber and A.W. Thomas,
\newblock Phys. Lett. B494 (2000) 89, hep-lat/0008018;
D.B. Leinweber, A.W. Thomas and R.D. Young,
\newblock Phys. Rev. Lett. 86 (2001) 5011, hep-ph/0101211;
T.R. Hemmert and W. Weise,
\newblock hep-lat/0204005.

\bibitem{MASSES}
D.B. Leinweber et~al.,
\newblock Phys. Rev. D61 (2000) 074502, hep-lat/9906027;
D.B. Leinweber, A.W. Thomas and S.V. Wright,
\newblock Phys. Lett. B482 (2000) 109, hep-lat/0001007;
D.B. Leinweber et~al.,
\newblock Phys. Rev. D64 (2001) 094502, arXiv:hep-lat/0104013;
S.V. Wright et~al.,
\newblock Nucl. Phys. Proc. Suppl. 109 (2002) 50, hep-lat/0111053.

\bibitem{STRUCTURE}
W. Detmold et~al.,
\newblock Phys. Rev. Lett. 87 (2001) 172001, hep-lat/0103006;
W. Detmold, W. Melnitchouk and A.W. Thomas,
\newblock Phys. Rev. D66 (2002) 054501, hep-lat/0206001.

\bibitem{STRANGE}
D.B. Leinweber and A.W. Thomas,
\newblock Phys. Rev. D62 (2000) 074505, hep-lat/9912052;
R. Lewis, W. Wilcox and R.M. Woloshyn,
\newblock hep-ph/0210064.

\bibitem{Guo:2001ph}
X.H. Guo and A.W. Thomas,
\newblock Phys. Rev. D65 (2002) 074019, hep-ph/0112040.

\bibitem{QUENCH}
R.D. Young et~al.,
\newblock Phys. Rev. D66 (2002) 094507, hep-lat/0205017;
R.D. Young et~al.,
\newblock Proceedings of the Workshop on Physics
at the Japan Hadron Facility (JHF) (2002) 155, nucl-th/0211026.

\bibitem{Donoghue:1998bs}
J.F. Donoghue, B.R. Holstein and B. Borasoy,
\newblock Phys. Rev. D59 (1999) 036002, hep-ph/9804281.

\bibitem{Bernard:2002yk}
C. Bernard et~al.,
\newblock (2002), hep-lat/0209086.

\bibitem{Gell-Mann:1968rz}
M. Gell-Mann, R.J. Oakes and B. Renner,
\newblock Phys. Rev. 175 (1968) 2195.

\bibitem{Li:1971vr}
L.F. Li and H. Pagels,
\newblock Phys. Rev. Lett. 26 (1971) 1204.

\bibitem{Borasoy:2002jv}
B. Borasoy et~al.,
\newblock Phys. Rev. D66 (2002) 094020, hep-ph/0210092.

\bibitem{Gasser:1988rb}
J. Gasser, M.E. Sainio and A. Svarc,
\newblock Nucl. Phys. B307 (1988) 779.

\bibitem{Lebed:1994yu}
R.F. Lebed,
\newblock Nucl. Phys. B430 (1994) 295, hep-ph/9311234.

\bibitem{Banerjee:1996wz}
M.K. Banerjee and J. Milana,
\newblock Phys. Rev. D54 (1996) 5804, hep-ph/9508340.

\bibitem{SVW}
S.V. Wright,
\newblock PhD thesis, University of Adelaide, 2002.

\bibitem{Cloet:2002eg}
I.C. Cloet, D.B. Leinweber and A.W. Thomas,
\newblock Phys. Rev. C65 (2002) 062201, hep-ph/0203023.

\bibitem{AliKhan:2001tx}
CP-PACS, A. Ali~Khan et~al.,
\newblock Phys. Rev. D65 (2002) 054505, hep-lat/0105015.

\bibitem{Iwasaki:1985we}
Y. Iwasaki,
\newblock Nucl. Phys. B258 (1985) 141.

\bibitem{Allton:1998gi}
UKQCD, C.R. Allton et~al.,
\newblock Phys. Rev. D60 (1999) 034507, hep-lat/9808016.

\bibitem{Sommer:1994ce}
R. Sommer,
\newblock Nucl. Phys. B411 (1994) 839, hep-lat/9310022.

\bibitem{Edwards:1998xf}
R.G. Edwards, U.M. Heller and T.R. Klassen,
\newblock Nucl. Phys. B517 (1998) 377, hep-lat/9711003.

\bibitem{Becher:1999he}
T. Becher and H. Leutwyler,
\newblock Eur. Phys. J. C9 (1999) 643, hep-ph/9901384.

\bibitem{Bernard:2001av}
C. Bernard et~al.,
\newblock Phys. Rev. D64 (2001) 054506, hep-lat/0104002.

\bibitem{McGovern:1998tm}
J.A. McGovern and M.C. Birse,
\newblock Phys. Lett. B446 (1999) 300, hep-ph/9807384.

\end{thebibliography}

\end{document}